\documentclass[conference]{IEEEtran}
\IEEEoverridecommandlockouts
\usepackage{cite}
\usepackage{amsmath,amssymb,amsfonts}
\usepackage{graphicx}
\usepackage{textcomp}
\usepackage{xcolor}
\usepackage{footnote}
\usepackage{algpseudocode}
\usepackage{algorithm}
\usepackage{subfig}
\usepackage{amsthm}
\usepackage{algorithmicx}
\algdef{SE}[DOWHILE]{Do}{doWhile}{\algorithmicdo}[1]{\algorithmicwhile\ #1}%

\def\BibTeX{{\rm B\kern-.05em{\sc i\kern-.025em b}\kern-.08em
    T\kern-.1667em\lower.7ex\hbox{E}\kern-.125emX}}
\begin{document}

\title{Extensions of a Line-Graph-Based Method for Token Routing in Decentralized Exchanges 
}

\author{\IEEEauthorblockN{Yu Zhang}
\IEEEauthorblockA{\textit{BDLT, IfI Department}, \\ \textit{University of Zurich}\\
Zurich, Switzerland 
\\Email: zhangyu@ifi.uzh.ch}
\and
\IEEEauthorblockN{Claudio J. Tessone}
\IEEEauthorblockA{\textit{BDLT, IfI Department}, \\ \textit{University of Zurich}\\
Zurich, Switzerland}
}

\maketitle
\begin{abstract}

Decentralized exchanges (DEXs) constitute a crucial component of the decentralized finance (DeFi) ecosystem, facilitating the daily trading of tokens worth billions of dollars. Nevertheless, a substantial proportion of these transactions are suboptimal, in that more target tokens could be acquired through alternative trading paths. This inefficiency underscores the urgency and importance of research into token routing, both from practical and academic perspectives. Zhang et al. (2025) proposed a linear line-graph-based approach tailored to scenarios involving a single DEX, demonstrating relevance not only in practical applications but also in theoretical research. However, actual trading environments in DEXs are considerably more complex, necessitating the development of extensions to the original line-graph-based methodology.
In this paper, we present three primary extensions to the original line-graph-based method. The first extension incorporates a breadth-first search (BFS) link iteration rule to reduce computational complexity and average execution time, while still preserving the method’s profitability. The second extension introduces route-splitting, wherein a large trade is divided into several smaller trades of equal size to mitigate price slippage. This approach generally yields higher average profits for traders, albeit at the cost of increased computational complexity. The third extension generalizes the original method—which was limited to a single DEX—to accommodate the more complex setting of a DEX aggregator. These three foundational extensions also serve as a basis for deriving additional enhancements.
Through empirical analysis using real pool reserve data from Uniswap V2 and Sushiswap V2, we demonstrate that all three extensions remain effective, both in terms of computational efficiency and in identifying more profitable trading routes.

\end{abstract}


\section{Introduction}
The Bitcoin blockchain, launched in 2009, is one of the most successful platforms supporting cryptocurrency payments. Following its introduction, hundreds of new blockchain platforms have emerged, among which Ethereum stands out for its support of smart contracts. A smart contract is essentially a self-executing piece of code deployed on the blockchain. This functionality enables users to define custom operations within blockchain ecosystems, including the creation of new tokens, token exchanges, staking, liquidation, borrowing and lending, the issuance of stablecoins, and more. As a result, many functions traditionally associated with conventional financial systems can now be implemented on smart-contract-enabled blockchains. After approximately a decade of development, this technological evolution has given rise to the domain of decentralized finance (DeFi). Within DeFi, decentralized exchanges (DEXs) play a critical role, facilitating token trading and exchange in a manner analogous to traditional foreign exchange markets.

Decentralized exchanges (DEXs) facilitate peer-to-peer token trading without the need for intermediaries, utilizing the Automated Market Maker (AMM) mechanism. Within a DEX, numerous token trading pairs—referred to as liquidity pools—are maintained through proportional token deposits provided by liquidity providers. Participation as a liquidity provider is open to any user. Traders can exchange one type of token for another by interacting with these liquidity pools, where the exchange rate is automatically determined by a predefined AMM function, such as the Constant Product Market Maker (CPMM) model. Under this model, the product of the reserves of the two tokens in the pool remains constant before and after a trade.
In contrast to traditional centralized foreign exchange markets—where the exchange rate is effectively constant due to substantial liquidity depth—DEXs are subject to price slippage, a critical factor in the optimization of trading routes. This slippage arises primarily from two factors: the comparatively shallow liquidity of DEX pools and the mechanics of the AMM model, such as the CPMM rule, which inherently links price to the reserve ratio of the tokens in the pool.

In decentralized exchanges (DEXs), multiple trading paths may exist for exchanging a given pair of tokens, potentially resulting in price discrepancies and subsequent arbitrage opportunities. While some studies have examined arbitrage mechanisms within DEXs, the primary function of these platforms remains the facilitation of token exchanges—that is, converting one type of token into another. The design of algorithms capable of identifying optimal trading paths is therefore a topic of considerable importance, both for practitioners seeking to maximize returns and for researchers investigating algorithmic efficiency and market behavior. An optimal trading path is defined as a sequence of trades that yields the maximum possible quantity of target tokens in exchange for a given quantity of source tokens.
To solve this problem, \cite{zhang2025line} proposed a line-graph-based method designed for the case of a single DEX, specifically targeting the scenario of linear routing. Linear routing refers to the process of token swapping through a sequential series of liquidity pools, where each trade follows directly from the previous one in a single path. For example, the path $A\rightarrow B\rightarrow C$ is a linear routing for trading token $A$ to token $C$. If half number of tokens $A$ is traded by the path $A\rightarrow B\rightarrow C$ and the rest half is traded by the path $A\rightarrow D\rightarrow C$, then the trading path is not a linear routing. There are also other trading cases that the paper \cite{zhang2025line} did not consider, such as DEX aggregators. In this paper, we will extend the line-graph-based method in \cite{zhang2025line} into different scenarios to make it more applicable in practice.

Among all decentralized exchanges, Uniswap—utilizing the Constant Product Market Maker (CPMM) model—is one of the most prominent platforms, with over 300,000 tokens created to date. Its daily trading volume reaches several billion dollars, solidifying its position as a leading DEX in the DeFi ecosystem. In this study, the majority of our analysis is based on data from Uniswap V2. For scenarios involving DEX aggregators, and for the sake of simplicity, we incorporate data from both Uniswap V2 and Sushiswap V2, as the latter also employs an AMM mechanism similar to that of Uniswap V2. The proposed method is, in principle, extensible to other DEXs operating under different AMM models.

The remainder of this paper is organized as follows: Section 2 presents a review of related work on token routing in decentralized exchanges (DEXs). Section 3 describes the pool reserve data utilized in this study. Section 4 provides an overview of the line-graph-based method, including its core logic in the context of token routing. In Section 5, we introduce several methodological extensions and report the results of corresponding experiments. Finally, Section 6 concludes the paper by summarizing key findings, discussing limitations, and outlining potential directions for future research.

\section{Related Work} 

A number of studies have already explored token routing in decentralized exchanges (DEXs). \cite{danos2021global} proposed a convex optimization-based framework to address both arbitrage and trade routing problems in DEXs. Building on this, \cite{angeris2022optimal} systematically developed an optimization-based approach for solving the arbitrage maximization problem, which can also be extended to trade routing scenarios in platforms such as Uniswap V2. More recently, \cite{chitra2025optimal} extended the convex optimization-based routing method introduced by \cite{angeris2022optimal} to accommodate Uniswap V4, incorporating its new functionality, such as hooks.

However, as noted by \cite{zhang2025line}, the convex optimization-based routing method suffers from high computational complexity. Moreover, the resulting token routing paths are often impractical for real-world applications. This is largely due to the fact that, as demonstrated in an experimental appendix of their work, the optimal paths frequently involve nearly every available liquidity pool, rendering implementation infeasible in a real-time trading environment.

In practice, decentralized exchanges (DEXs) such as Uniswap primarily employ the depth-first search (DFS) algorithm to identify token routing paths. While the DFS algorithm is characterized by low computational complexity, the profitability of the routing paths it produces is typically limited, as noted by \cite{zhang2025line}. In the context of trading a source token for a target token, the profitability of a given path refers to the quantity of target tokens received in exchange for a fixed amount of source tokens.
To address the respective strengths and limitations of existing token routing algorithms, \cite{zhang2025line} proposed the line-graph-based routing method. This approach achieves a higher level of profitability than DFS, though it remains less profitable than the convex optimization-based method. In terms of computational complexity, it occupies a middle ground—lower than the convex optimization-based method, but higher than DFS. This analysis highlights that the line-graph-based method represents a trade-off between computational efficiency and profitability, balancing the extremes of the DFS and convex optimization-based approaches.

\section{Data Description}

This study utilizes pool reserve data from Uniswap V2 and Sushiswap V2. Sushiswap V2 applied the same AMM function as Uniswap V2 in token trading.
Based on the liquidity pools from a DEX, a token graph is constructed wherein nodes represent individual tokens and edges correspond to liquidity pools that contain the respective tokens at the endpoints. Additional details regarding the criteria for token filtering during the construction of the token graph can be found in \cite{zhang2024improved}. In the experimental evaluations of each extension to the line-graph-based method, the resulting graphs comprise approximately one hundred tokens and four hundred directed edges.

\section{Review of the Line-graph-based Routing Method}

In this section, we briefly review the line-graph-based linear routing method developed in \cite{zhang2025line} under the case of a single DEX.

Firstly, the tokens' liquidity pool data was used to construct the directed token graph $G(V, E, R)$  (or $G$ for short) with $V$ being the token set, with $E$ being the token liquidity pool set, and with $R$ being liquidity pools' token reserve sets in the form of $\{(v_i,v_j):(r_i,r_j), \cdots\}$, $i\neq j$. Each directed edge $(v_i,v_j)$ denotes an unidirectional liquidity pool, meaning users can only trade $v_i$ for $v_j$, but not vice versa. After the token graph is built, some token filtering rules are used to select liquidity pools with deeper liquidity, which can be referred to \cite{zhang2024improved}.

Secondly, the line graph $L(G)$ of the token graph $G(V, E, R)$ is constructed. Each directed edge will be a new vertex in the line graph $L(G)$ with corresponding liquidity reserve information stored. For example, the directed edge $(v_i,v_j)$ from $G$ will be the vertex $(v_i,v_j)$ in the $L(G)$ with the corresponding unidirectional pool reserve information $(v_i,v_j):(r_i,r_j)$ stored in the vertex $(v_i,v_j)$ of the line graph $L(G)$.
The directed links from a vertex to all other vertices can be built under the condition that the last token of the specified vertex is exactly the first token of the other corresponding vertices. For example, $(v_i, v_j)$ can connect to $(v_j,v_l)$. To simplify the calculation, the authors in \cite{zhang2025line} cut redundant vertex links in the line graph and added an extra vertex that connects to all its neighbour vertices. Assuming the added vertex is $v_1$, called the source token, then all vertices with the first token being $v_1$ are the extra vertex $v_1$'s neighbour vertices.

Each vertex in $L(G)$ also stores the trading path from the source token $v_1$ to the target token which is the second token in the vertex $(v_i,v_j)$, and the number of target tokens obtained in the corresponding paths. In the vertex $(v_i,v_j)$, $v_j$ is an target token. All target tokens' paths are initialized as null, and the corresponding numbers of tokens obtained are set to zero.

The algorithm starts from the extra vertex ($v_1$) with inputting $\epsilon$ units of source token $v_1$. 
In the first round of iteration, each directed vertex link in $L(G)$ is randomly selected and then iterated to calculate the maximal output of another target token for a given input of another token, which is stored in the initial vertex of the link. For example, in the link $(v_i,v_j)\rightarrow(v_j,v_k)$ from $L(G)$, the number of $v_k$ obtained can be calculated by selling $v_j$ with $v_j$'s inputting count stored in $(v_i,v_j)$. If the iteration of a directed vertex link results in a greater quantity of target tokens at the terminal vertex, the corresponding trading path is updated to consist of the trading path associated with the initial vertex of the iterated link concatenated with the terminal vertex of that link. Concurrently, the target token quantity at the terminal vertex is recalculated and updated accordingly.

Once all links have been iterated over in the first round, the second round of interactions commences, and so on.
The link iteration stops until all target tokens' counts stored in each vertex of $L(G)$ do not increase. The detailed introduction of the line-graph-based token routing algorithm can be found in \cite{zhang2025line}.

The paper found that users can obtain more target tokens with this line-graph-based method, compared to the depth-first-search (DFS) used by Uniswap V2, while the trading path lengths from the two methods are comparable.

\section{Extensions of the Line-graph-based Method}
\subsection{Breadth First Search (BFS) Traversal Rule in Line-graph-based Method in A Single DEX}\label{bfs_section}

After the line graph of the token graph is constructed, the vertex link iteration starts. In each round of iteration in the line-graph-based method in \cite{zhang2025line}, the iterated link is selected randomly each time, which may make the iteration process inefficient. For example, assuming the link $(v_i,v_j)\rightarrow(v_j,v_k)$ is selected to iterate, if $v_j$'s count on $(v_i,v_j)$ is still zero, then information on $(v_j,v_k)$ will not be updated, leading to more rounds of iteration. Based on this analysis, we put forward the breadth-first-search (BFS) link iteration rule that defines the link iteration order in each round of link iteration to increase the link iteration efficiency and decrease the token routing calculation time. 

By using the BFS link iteration rule, we can ensure that each link iteration is an effective iteration, namely, the input token's count in trading is non-zero. In the directed line graph $L(G)$, the defined BFS link interaction order algorithm is as follows.

\begin{algorithm}[H]
    \caption{BFS Link Iteration Order Algorithm}\label{Algorithm 2}
\textbf{Problem setting:} \textit{Return the link iteration order for the line graph $L(G)$.}

    \hspace*{\algorithmicindent} \textbf{Input}: source token $src$, line graph $L(G)$\\
    \hspace*{\algorithmicindent} \textbf{Output}: $edge_{order}$, link iteration order list.
    
\begin{algorithmic}[0]
    \State $queue$ = [$src$]. Iteration starting from the source token $src$.
    \Do
        \For{each $q$ in $queue$}
            \State $neigh$ = $q$'s neighbour in $L(G)$
            \If{$neigh$ is not empty}
                \For{each $n$ in $neigh$}
                    \If{vertex $(v,n)$ not in $edge_{order}$}
                        \State $edge_{order}.append((v,n))$
                        \State remove edge $(v,n)$ from $L(G)$
                    \EndIf
                \EndFor
            \EndIf
        \EndFor
    \doWhile{$L(G)$ has edges}
    
    \Return $edge_{order}$.
\end{algorithmic}
\end{algorithm}

\begin{figure}[htbp]
    \centering
    \includegraphics[width=0.9\linewidth]{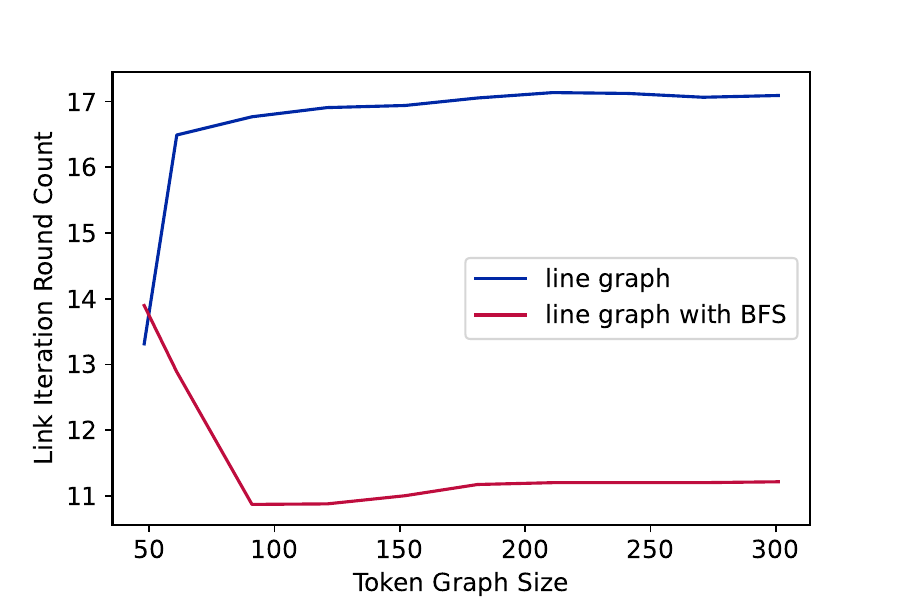}
    \caption{Average round of vertex link iteration. The blue and red lines denote the average rounds of vertex link iteration by using the original line-graph-based routing method in \cite{zhang2025line} and the extended method with BFS vertex link iteration, respectively.}
    \label{loop_count}
\end{figure}

To compare the effect of this extension to the original line-graph-based method in \cite{zhang2025line}, we calculate the average rounds of link iterations of both versions of the line-graph-based token routing methods in token graphs with different sizes. The result is shown in Fig. \ref{loop_count}. 

Fig. \ref{loop_count} shows that the number of link-iteration rounds indeed decreases a lot by using the line-graph-based method with the BFS link iteration rule compared to the original line-graph-based routing method in \cite{zhang2025line}. For example, in a token graph with around one hundred tokens, the average number of link-iteration rounds using the original line-graph-based method is 17, while the number is 11 by using the BFS iteration rule, which is a large increase in calculation efficiency.

\begin{figure}
\centering
\subfloat[]{\includegraphics[width=0.9\linewidth]{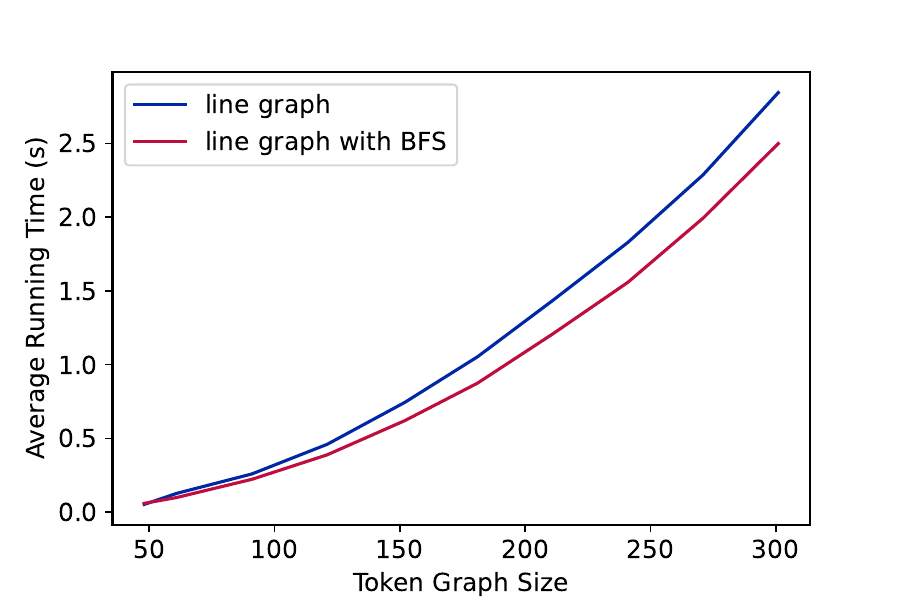}}

\centering
\subfloat[]{\includegraphics[width=0.9\linewidth]{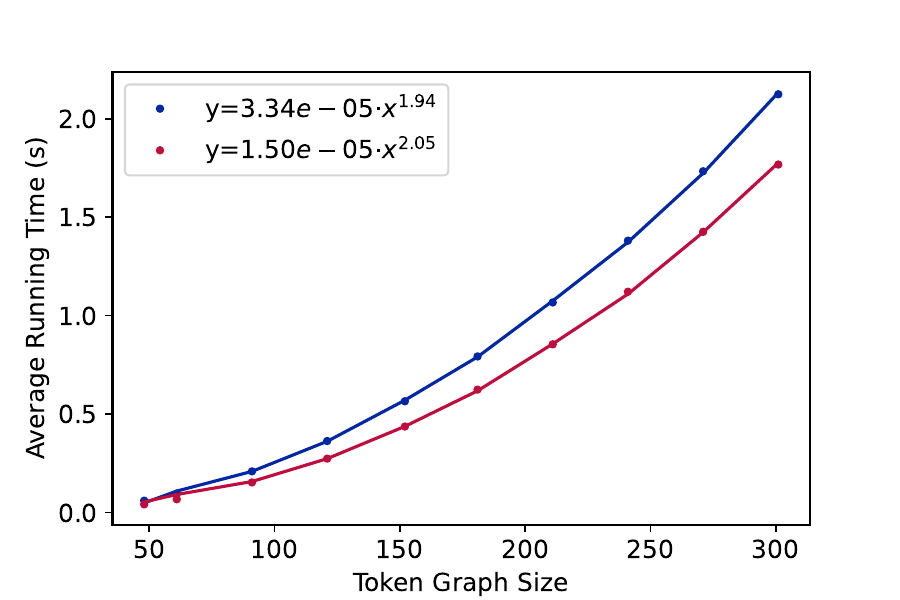}}
\caption{Average running time of the line-graph-based methods. The blue and red lines denote the average running time of the original line-graph-based routing method in \cite{zhang2025line} and the extended method with BFS vertex link iteration, respectively. In panel b, by fitting both lines, the average running time is almost a polynomial function of the size of the token graph.}
\label{average_running_time}
\end{figure}

\begin{figure}[htbp]
    \centering
    \includegraphics[width=0.9\linewidth]{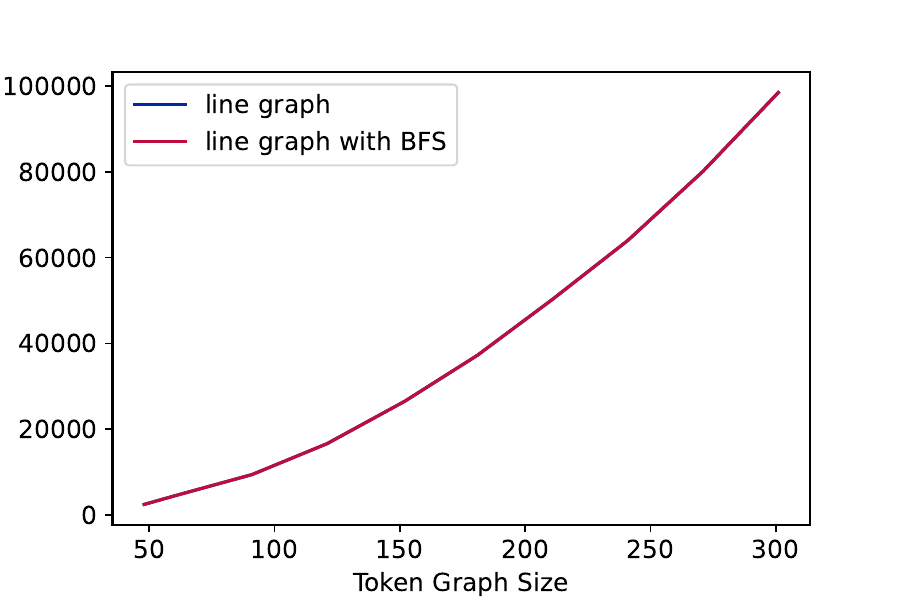}
    \caption{Relationship between the number of links in the line-graph and the size of the underlined token graph}
    \label{line_graph_edge_size}
\end{figure}

Fewer link iteration rounds will enhance computational efficiency and reduce computation time, which is shown in Fig. \ref{average_running_time}. In this figure, the red line denotes the average running time by using the line-graph-based method with BFS iteration rule. We can find that it is below the blue line, which denotes the average running time of the line-graph-based method in \cite{zhang2025line}. Through the average running time grows fast with the token graph size, the fitting result in panel b of Fig.\ref{average_running_time} shows that the computation complexity is a polynomial function of the graph size, which sources from the fact that the number of edges in the line graph of the original token graph increase almost in a polynomial function with the original token size, as shown in Fig. \ref{line_graph_edge_size}. As pointed out in \cite{zhang2025line},the number of vertices in the $L(G)$ ($M_{L(G)}$) equals the number of edges ($E_G$) in $G$, namely, $M_{L(G)}=E_G$. The number of links ($E_{L(G)}$) in $L(G)$  equals the sum of the degree's square of each node minus two times the number of edges in graph $G$, namely $E_{L(G)} =\sum {d_i}^2-2E_G$, where $d_i$ denotes the degree of token (node) $i$ in $G$ and $E_G$ denotes the number of edges in $G$. For graphs of substantial size, the corresponding line graph may contain a large number of links. Nevertheless, in token graphs comprising approximately one thousand nodes, like PancakeSwap, the line-graph-based method remains highly efficient and can serve as a valuable complement to the depth-first search (DFS) algorithm currently employed by Uniswap.

\begin{figure}[htbp]
    \centering
\subfloat[]{ \includegraphics[width=0.9\linewidth]{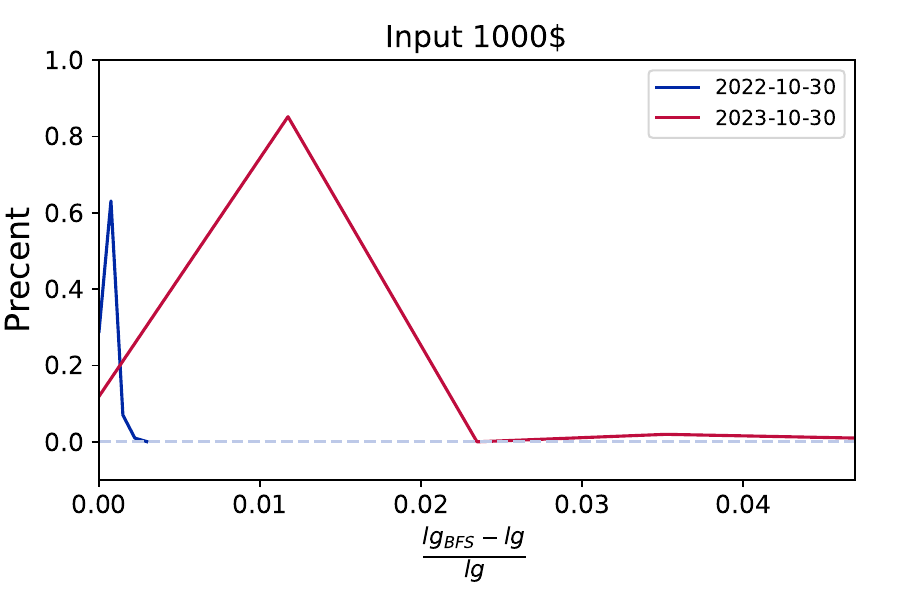}
    \label{bfs_1000}}
    
    \centering
\subfloat[]{
    \includegraphics[width=0.9\linewidth]{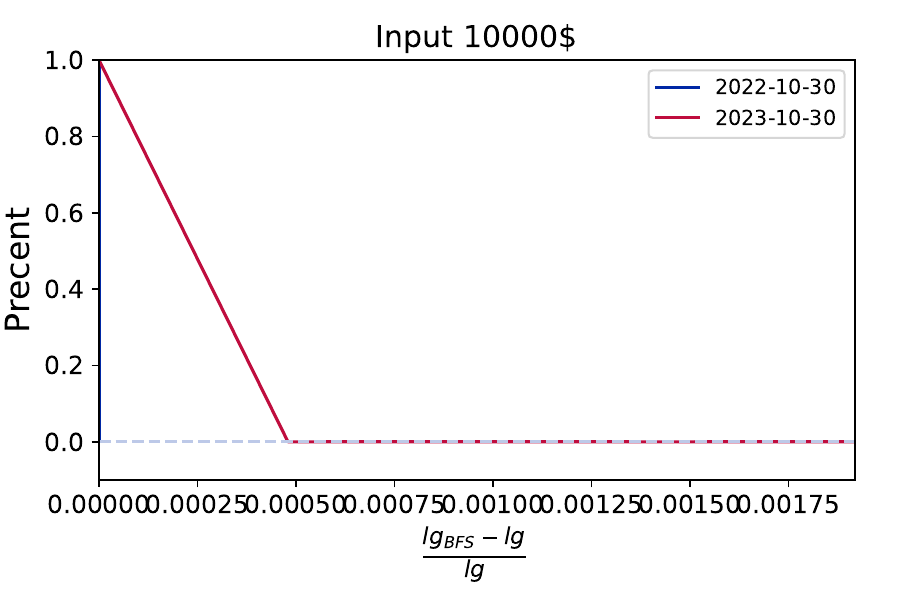}
    \label{bfs_10000}}
\caption{Profitability comparison between the line-graph-based method in \cite{zhang2025line} and that with BFS iteration rules. For any pair of source token and target token, $lg$ and $lg_{BFS}$ represent the number of target tokens produced by investing a specific number of source tokens, using the line-graph-based method described in \cite{zhang2025line} and the method employing BFS iteration rules, respectively. The input number of source tokens is calculated as: $\frac{M}{p}$, where $M$ denotes the amount of capital invested, and $p$ denotes a token's price. $M$ is 1000\$ and 10000\$, respectively, in panels a and b of the figure. The red and blue lines correspond to the analysis results using data on the 30th of October, in 2023 and 2022, respectively.}
\label{bfs_profitability}
\end{figure}

We also compared the profitability of the extended algorithm with the BFS link iteration rule to the original line-graph-based method in \cite{zhang2025line}, and the result is shown in Fig.\ref{bfs_profitability}. Fig.\ref{bfs_profitability} shows that the profitability of the line-graph-based method with BFS extension is a bit more profitable than the original line-graph-based method in \cite{zhang2025line}.

\subsection{Routes Splitting Algorithm in A Single DEX}

The token routing method proposed in \cite{zhang2025line} follows a linear routing approach. However, the constant product market maker (CPMM) exchange functions commonly used in decentralized exchanges (DEXs) inherently suffer from price slippage during trading. Specifically, larger trade sizes within a liquidity pool result in greater price slippage. To mitigate slippage and maximize the number of target tokens obtained when exchanging a fixed amount of source tokens, a natural strategy is to divide a large trade into multiple smaller trades executed sequentially. After a smaller trade is executed along the optimal trading path determined by the line-graph-based routing method, the reserve information of the relevant liquidity pools is updated, and the optimal path for the subsequent trade—of the same reduced size—is recalculated.

Based on this analysis, we put forward the routing splitting algorithm in the case of a single DEX. The exact algorithm is as follows:

\begin{algorithm}[hbpt]
    \caption{Routes Splitting Algorithm in A Single DEX}\label{Algorithm 2}
\textbf{Problem setting:} \textit{ Buy the maximal units of target token $v_N$ ($\Delta r_n$) with inputting $\epsilon_1$ units of source token $v_1$ in a single DEX.}

    \hspace*{\algorithmicindent} \textbf{Input}: line graph $L(G)$, $\epsilon_1$ units of source token $v_1$\\
    \hspace*{\algorithmicindent} \textbf{Output}: Maximal units of target token $v_n$
\begin{algorithmic}[0]    
    \State $K \gets$ the number of splits. 
    \State $v_n$ $\gets$ $0$
    \State $P$ $\gets$ the path set.
    \For{$k$ $<$ $K$}
        \State Target token count $T$, trading path $TP$ = 
        Line-graph-based algorithm in \cite{zhang2025line}($\frac{\epsilon_1}{K}$)
        \State $v_n += T$
        \State $P.add(TP)$
        \State Update the reserve of pools that are in the trading path.
    \EndFor
    
    \Return $v_n, P$
\end{algorithmic}
Reorder $P$ to make it an implementable path.
\end{algorithm}

Compared to the basic line-graph-based method proposed in \cite{zhang2025line}, the aforementioned trade-splitting algorithm—under otherwise identical settings—inevitably results in increased computational complexity. Specifically, the computational complexity grows linearly with the number of trade splits. For instance, if the trade is divided into two parts, the line-graph-based method must be executed once to determine the initial optimal path, the reserve information in the corresponding liquidity pools must then be updated, and the method must be executed a second time to determine the new optimal path. As this computational complexity analysis is straightforward and self-evident, we do not include a separate figure to illustrate it.

\begin{figure}[htbp]
\centering
\subfloat[]{ \includegraphics[width=0.9\linewidth]{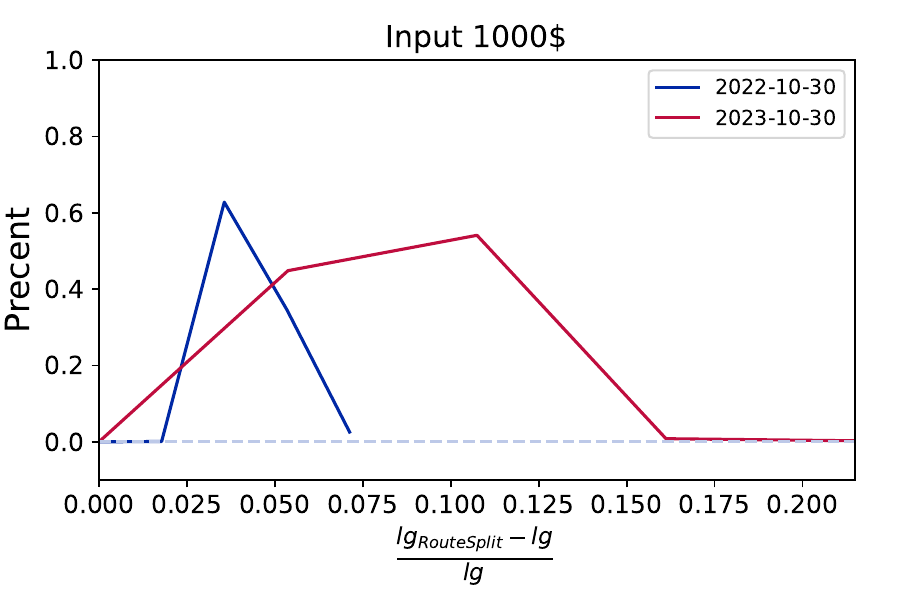}
    \label{routesplit_1000}}

    \centering
\subfloat[]{
    \includegraphics[width=0.9\linewidth]{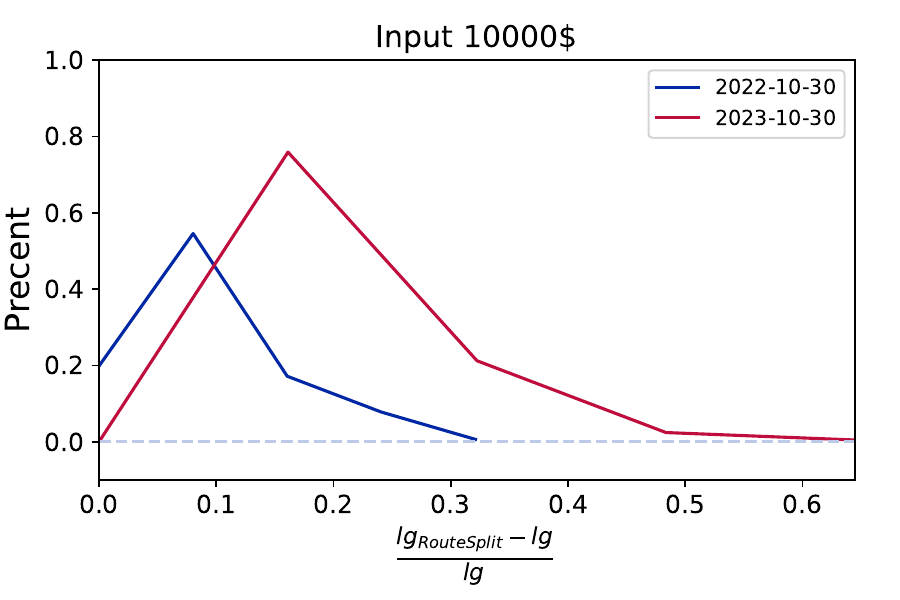}
    \label{routesplit_10000}}
    \caption{Profitability comparison between the line-graph-based method and the route splitting method. $lg$ and $lg_{RouteSplit}$ 
    represent the number of target tokens produced by investing a specific number of source tokens, using the line-graph-based method described in \cite{zhang2025line} and the method using route splitting, respectively.
    The explanations of other signs are the same as those in Fig.\ref{bfs_profitability}'s caption.}
    
\label{route_split_profitability}
\end{figure}

The primary objective of route splitting is to enhance trading profitability. To evaluate its effectiveness, we quantified the extent to which the route splitting method improves traders' profits in comparison to the original line-graph-based method presented in \cite{zhang2025line}. The results are presented in Fig. \ref{route_split_profitability}.

As shown in Fig.\ref{route_split_profitability}, traders' profits improved a lot by using the route splitting method compared to the original line-graph-based method in \cite{zhang2025line}. For example, with the data from Uniswap V2 on the 30th of October, 2023, if we input 10000\$, traders' profits can increase by more than 20\% in around 20\% cases, which is shown as the red line in panel (b).

\subsection{Linear Routing Algorithm in DEX Aggregators and Breadth First Search (BFS) Traversal Rules to Speed up}

The line-graph-based token routing method proposed in \cite{zhang2025line}, along with the two extensions presented earlier in this paper, are designed for scenarios involving a single decentralized exchange (DEX). However, the Ethereum blockchain currently hosts a wide range of DEXs, and traders frequently execute transactions across multiple platforms. In multi-DEX trading scenarios—such as those handled by DEX aggregators—the original line-graph-based algorithm and its aforementioned extensions are not directly applicable. In this section, we extend the line-graph-based algorithm to accommodate the DEX aggregator setting, focusing exclusively on the linear routing case.

A key distinction between the single-DEX scenario and that of a DEX aggregator (involving multiple DEXs) is the presence of multiple liquidity pools for the same token pair. When constructing the line graph in the context of a DEX aggregator, it is essential to preserve all possible combinations of liquidity pools corresponding to each token pair. Once the line graph is constructed, the vertex link iteration procedure closely follows the steps outlined in the original line-graph-based token routing method proposed in \cite{zhang2025line}. In the following part, we will mainly focus on how to construct the line graph for a DEX aggregator.

Consider an aggregator comprising two decentralized exchanges, denoted as $DEX_1$ and $DEX_2$. We assume that $DEX_2$ is either identical to or fully embedded within $DEX_1$, meaning that every token available on $DEX_2$ is also present on $DEX_1$.

We build the original token graph and the corresponding line graph for a DEX aggregator as follows:
\begin{enumerate}
    \item Firstly, we build the original token graph $G^1(V^1,E^1,R^1)$ and $G^2$ for $DEX_1$ and $DEX_2$, respectively, just like that in \cite{zhang2025line}. The only difference is that we add a superscript for each node, edge (liquidity pool), and reserve information, which denotes that the nodes and pools with reserve information are from $DEX_1$.    
    For example, the newly constructed graph is $G(V^1, E^1, R^1)$. The edge from token $v_i$ to $v_j$ is denoted as $e_{ij}^1$ or $(v_i,v_j)^1$ with reserve information denoted as $(v_i,v_j)^1=(r_i,r_j)^1$.

    \item Secondly, we build the line graph $L(G^1)^1$ for $G(V^1, E^1, R^1)$ and line graph $L(G^2)^2$ with the same steps as in \cite{zhang2025line}. One difference is that we need to add superscripts for the new vertices and reserve information. For example, The token reserve information of vertex $(v_i, v_j)^1$ in $L(G^1)^1$ is set as $(r_j, r_l)^1$.

    \item Thirdly, we integrate the two line graphs ($L(G^1)^1$, $L(G^2)^2$) as a integrated bigger line graph $L(G^1,G^2)$ with following steps:
        \begin{enumerate}
            \item Iterating each new vertex $(v_i,v_j)^1$ in $L(G^1)^1$. If there is a new vertex $(v_l,v_k)^2$ in $L(G^2)^2$ with $l=j$, then building a new link from  $(v_i,v_j)^1$ to $(v_l,v_k)^2$ which is denoted as $(v_i^1,v_j^1,v_k^2)$ or $(v_i,v_j)^1\rightarrow (v_j,v_k)^2$

            \item Iterating each new vertex in $L(G^2)^2$ and building new links with the same procedures as above.

            \item Adding an extra node and connecting it to all its neighbouring vertices. For example, if the added extra node is $v_1$, then link it to all vertices whose first token is $v_1$.
        \end{enumerate}
\end{enumerate}

Once the integrated line graph for the DEX aggregator is constructed, the vertex link iteration proceeds in the same manner as described in the line-graph-based method in \cite{zhang2025line}, and is therefore not repeated here.

In the case of a DEX aggregator comprising two distinct decentralized exchanges, liquidity pool data is sourced from both Uniswap V2 and Sushiswap V2. When constructing the token graph, all tokens from Sushiswap V2 are assumed to be included within Uniswap V2. As a result, the integrated line graph representing the DEX aggregator contains a greater number of vertex links compared to that of a single DEX.

Suppose there are $n$ distinct line graphs, each possessing $E_{L(G)}$ links and sharing the same connection structure. In this case, the total number of links in the integrated line graph is $n^2 \cdot E_{L(G)}$. Consequently, the computational complexity associated with the DEX aggregator scenario is higher than that of a single-DEX setting.

\begin{figure}[htbp]
\centering
\subfloat[]{ \includegraphics[width=0.9\linewidth]{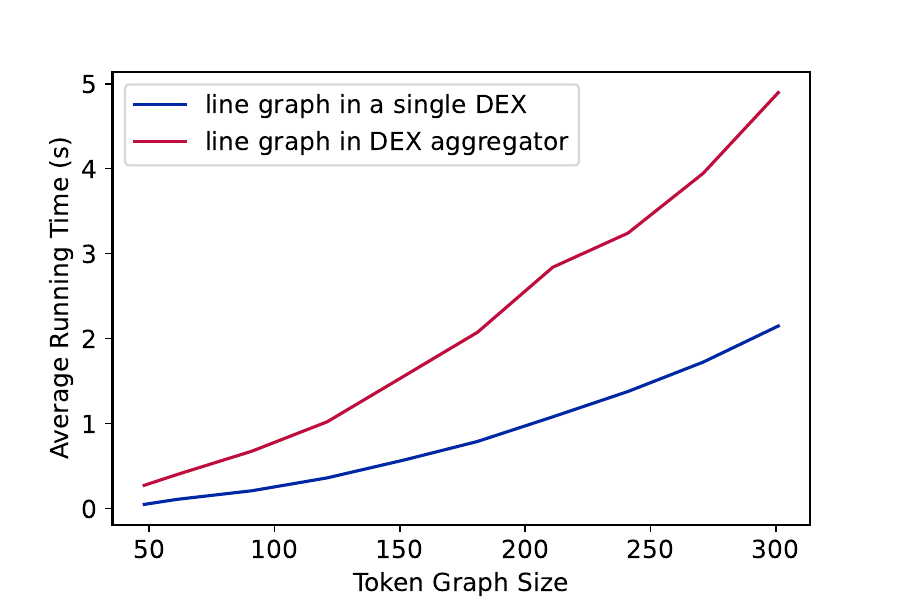}
    \label{multi_dex_dfs_1000}}
    
    \centering
\subfloat[]{
    \includegraphics[width=0.9\linewidth]{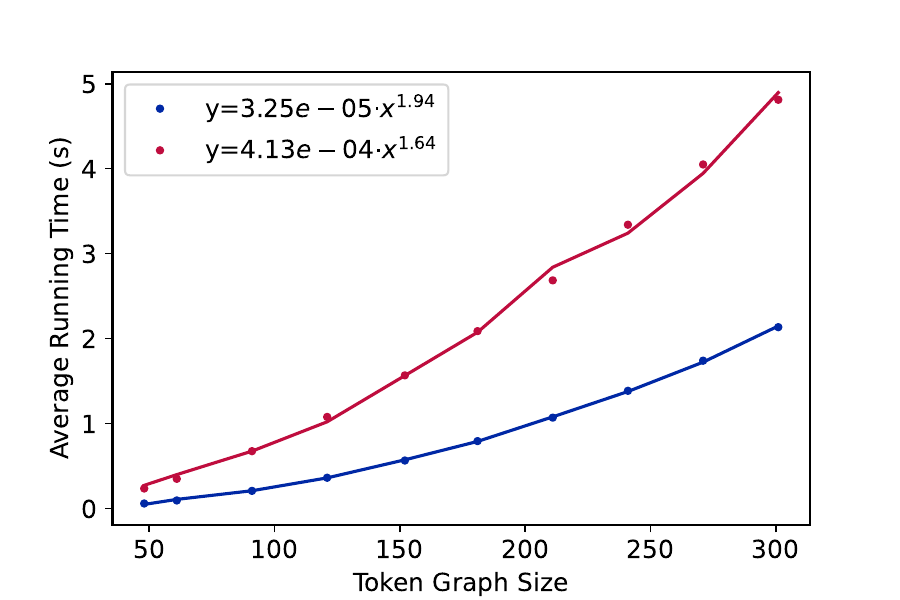}
    \label{multi_dex_dfs_10000}}
\caption{Average running time of the line-graph-based routing method in DEX aggregator and a single DEX. The red and blue lines denote the average running time in a DEX aggregator and a single DEX, respectively, by running the line-graph-based method. In panel b, we find that the polynomial function fits very well to the size of the token graph.}
\label{average_running_cost_multi_dex}
\end{figure}

In Fig.\ref{average_running_cost_multi_dex}, we calculate the average running time of each token routing method. We can find that the average running time by running the line-graph-based method in a DEX aggregator (Unisawap V2 + Sushiswap V2) is clearly higher than running the same algorithm in a single DEX (Unisawap V2). By fitting, the line average running time is almost a polynomial function of the token graph size, as shown in panel b of Fig. \ref{average_running_cost_multi_dex}.

\begin{figure}[htbp]
\centering
\subfloat[]{ \includegraphics[width=0.9\linewidth]{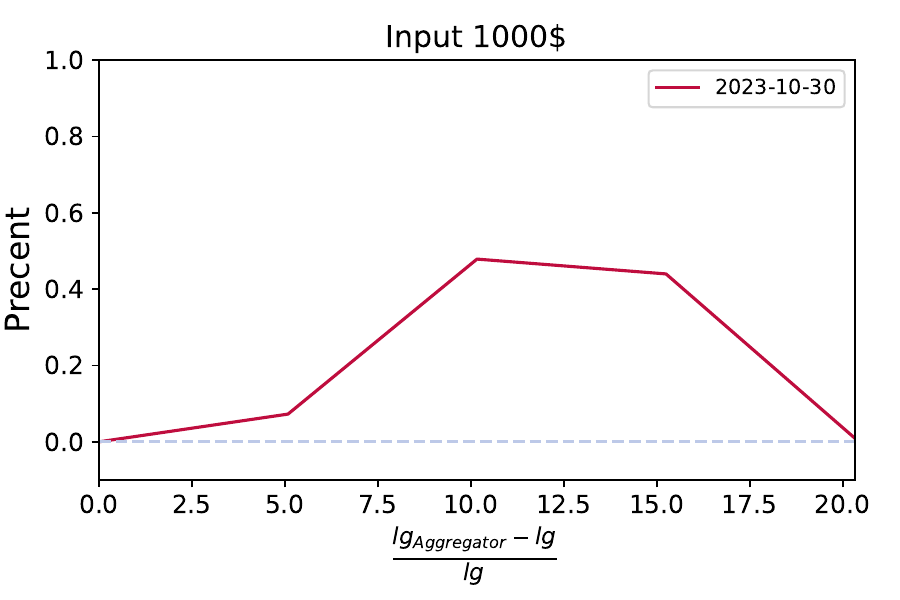}
    \label{multi_dex_lg_1000}}
    
\centering
\subfloat[]{
    \includegraphics[width=0.9\linewidth]{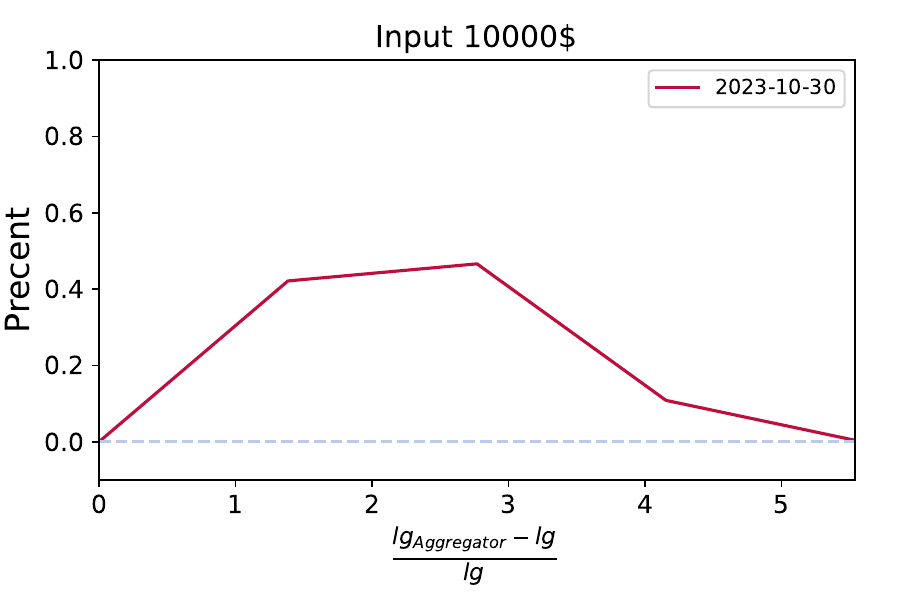}
    \label{multi_dex_lg_10000}}

\caption{The profitability performance comparison between the scenario of DEX aggregator and a single DEX by using the line-graph-based method. $lg_{Aggregator}$ denotes the profitability of the line-graph-based method in a DEX aggregator with vertex link randomly selected during iteration, and $lg$ denotes the profitability by running the same routing method in a single DEX with the vertex link in each round randomly selected, just as in \cite{zhang2025line}. The blue and red lines denote that the experiment was conducted on data from 2022 and 2023, respectively.}
\label{Profitablity_multi_dex_vs_single_dex}
\end{figure}

Then we compare the profitability of the line-graph-based method in the scenario of a DEX aggregator and of a single DEX, which is shown in Fig.\ref{Profitablity_multi_dex_vs_single_dex}. From this figure, we can find that the profitability of the line-graph-based method in a DEX aggregator is much higher than that in a single DEX, which is easy to understand because more profitable trading paths are included in a DEX aggregator than in only a single DEX from the DEX aggregator.

We can find that the trading path with much higher profit can be detected in a DEX aggregator than in a single DEX.

In the line graph of a DEX aggregator, the DFS token routing algorithm can usually detect a more profitable path than in a single DEX. The logic of the DFS algorithm in a DEX aggregator is similar to that in a single DEX, which has been described in \cite{zhang2025line}, and we didn't repeat it in this paper.

\begin{figure}[htbp]
\centering
\subfloat[]{ \includegraphics[width=0.9\linewidth]{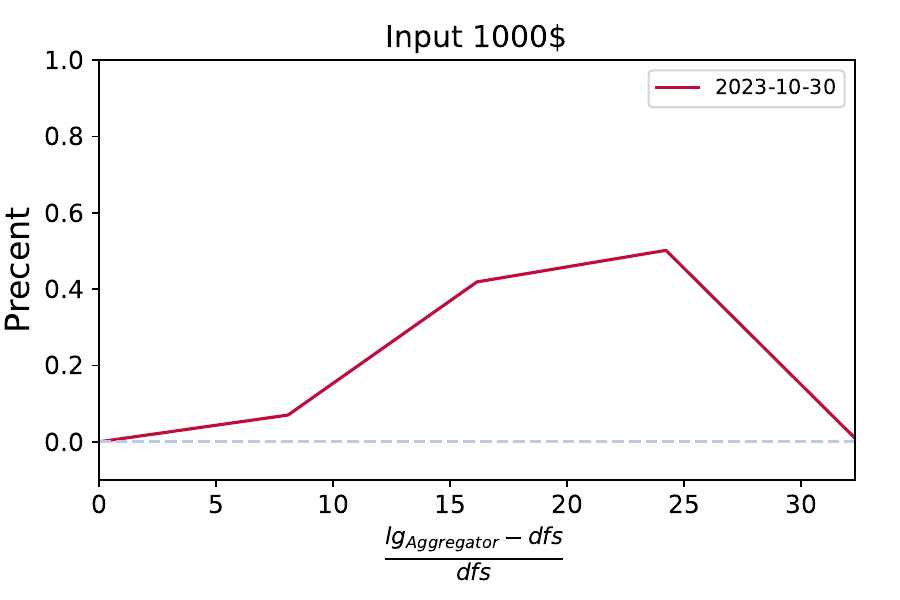}
    \label{multi_dex_dfs_1000}}

\centering
\subfloat[]{
    \includegraphics[width=0.9\linewidth]{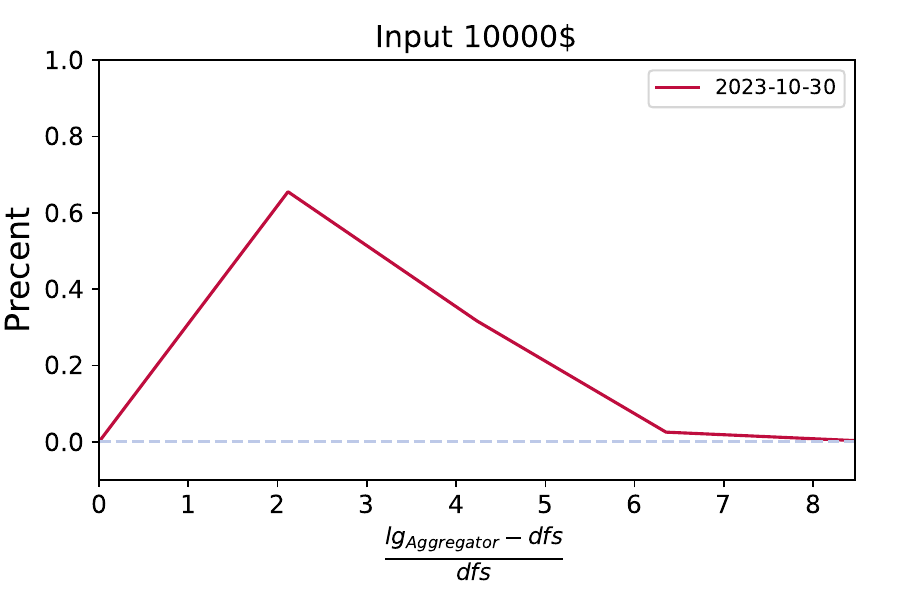}
    \label{multi_dex_dfs_10000}}
\caption{A profitability comparison between line-graph-based routing methods in a DEX aggregator and the DFS algorithm in the same DEX aggregator. $lg_{Aggregator}$ denotes the profitability of the line-graph-based method, and $dfs$ denotes the profitability by running the DFS algorithm in the corresponding token graph.}
\label{multi_dex_dfs}
\end{figure}

However, we are not sure about the profitability performance of this extension of the line-graph-based method in a DEX aggregator compared to the DFS routing algorithm in the same setting.
In the following experiment, we compare their profitability of the line-graph-based method and the DFS algorithm in a DEX aggregator. The result is shown in Fig. \ref{multi_dex_dfs}.

As shown in Fig.\ref{multi_dex_dfs}, the line-graph-based method in a DEX aggregator is still much more profitable than the DFS algorithm in the same corresponding token graph. For example, based on the token reserve information in 2023, if we input 10000\$, then in around 60\% cases, the line-graph-based method in a DEX aggregator can outperform the DFS algorithm around 5 times.

In the extension presented in this subsection, once the line graph of the DEX aggregator is constructed, vertex links are selected randomly during each iteration round. As demonstrated in Section \ref{bfs_section}, employing the breadth-first search (BFS) iteration rule can accelerate the line-graph-based method, resulting in reduced average token routing time. However, the impact of the BFS iteration rule on profitability, in comparison to the random link traversal strategy, remains uncertain in the context of this extension.

\begin{figure}[htbp]
\centering
\subfloat[]{ \includegraphics[width=0.9\linewidth]{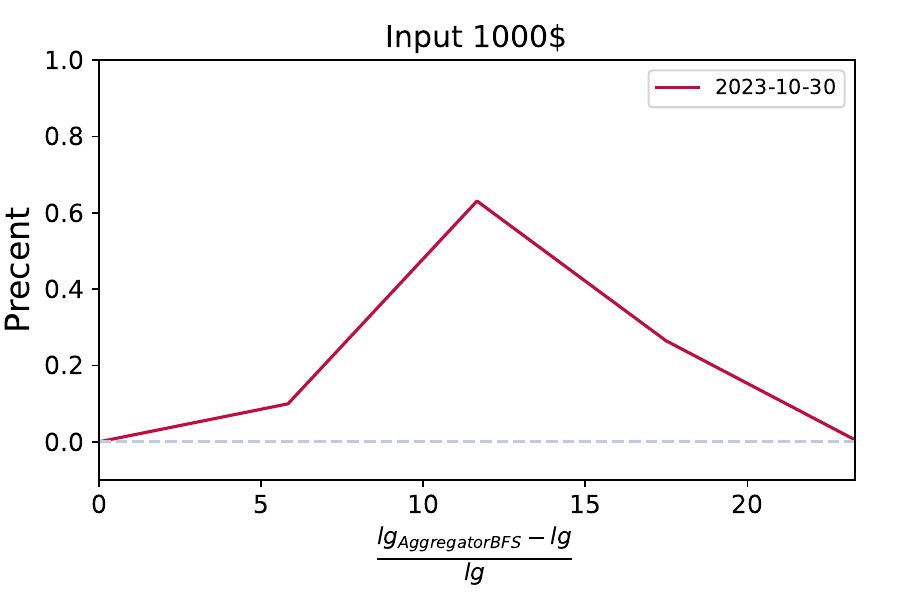}}
    \label{multi_dex_dfs_1000}

\centering
\subfloat[]{
    \includegraphics[width=0.9\linewidth]{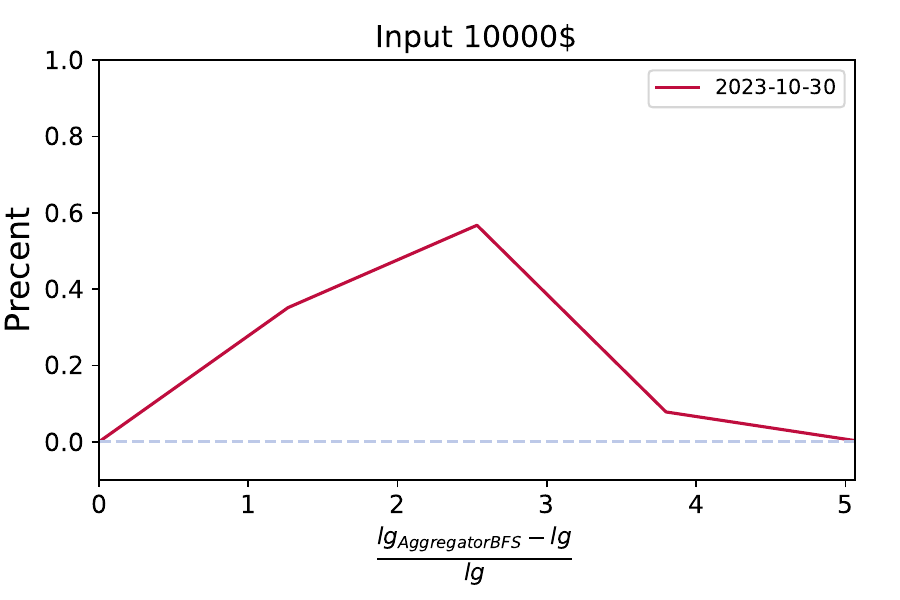}
    \label{multi_dex_dfs_10000}}
\caption{A profitability comparison between line-graph-based routing methods in a DEX aggregator using BFS versus random link traversal rules. $lg_{AggregatorBFS}$ denotes the profitability of the line-graph-based method with BFS link iteration rule in a DEX aggregator, and $lg$ denotes the profitability by running the same routing method in a single DEX.}
\label{multi_dex_with_bfs}
\end{figure}

To evaluate this, we conduct an experiment comparing the profitability of the two line-graph-based methods—one using the BFS iteration rule and the other using random link traversal. The results are presented in Fig. \ref{multi_dex_with_bfs}.

As shown in Fig. \ref{multi_dex_with_bfs}, the line-graph-based routing method with BFS vertex link iteration rule in a DEX aggregator is still much profitable than the line-graph-based routing method with a random link iteration rule in a single DEX.

\section{Results and Discussions}
This paper primarily investigates three key extensions to the line-graph-based linear token routing method within the context of a single decentralized exchange (DEX). These extensions include: (1) incorporating a breadth-first search (BFS) link iteration rule to accelerate the computation of the linear line-graph-based routing algorithm; (2) splitting large token trades into smaller transactions to enhance overall trading profitability; and (3) adapting the linear line-graph-based routing method to the multi-DEX setting of a DEX aggregator.

Based on experiments conducted using Uniswap V2 data, we observed that the incorporation of the BFS link iteration rule significantly improves the efficiency of the line-graph-based token routing method by enhancing token iteration performance. The average runtime exhibits approximately polynomial growth with respect to the size of the original token graph, which is attributable to the fact that the number of links in the line graph also scales polynomially with the size of the original token graph. Furthermore, our results indicate that the line-graph-based method employing the BFS iteration rule can enhance trading profitability compared to the original method in \cite{zhang2025line}, which relies on random link iteration.

The increased price slippage associated with large trade sizes in decentralized exchanges (DEXs) motivates the extension of our line-graph-based method to incorporate route splitting. In this extension, a large trade is evenly divided into smaller trade sizes, each of which is sequentially executed to identify an optimal trading path. Compared to the original linear line-graph-based method proposed in \cite{zhang2025line}, the route-splitting approach offers a substantial improvement in trading profitability.

The token routing method presented in \cite{zhang2025line}, along with the two aforementioned extensions, is applicable only within the context of a single decentralized exchange (DEX). However, in practice, numerous DEXs coexist within the ecosystem. To address this, we extend the linear line-graph-based token routing method to accommodate a DEX aggregator setting. Experimental results based on pool reserve data from Uniswap V2 and Sushiswap V2 demonstrate that this extended method significantly enhances trading profitability compared to applying the linear line-graph-based method within a single DEX alone.

Numerous additional extensions are also possible, many of which involve various combinations of the three primary extensions discussed above. For instance, route splitting can be applied within the context of a DEX aggregator—an approach that is briefly outlined in the appendix.

Based on our analysis of computational complexity and profitability, the line-graph-based token routing method appears to strike a balance between the convex optimization approach and the depth-first search (DFS) algorithm.

Efficient token routing in decentralized exchanges is critical for traders, as it has a direct impact on trading outcomes and profitability. As such, this topic warrants further in-depth research.

\bibliographystyle{IEEEtran}
\bibliography{reference.bib}


\section{Appendix}

\subsection{Routes Splitting Algorithm in DEX Aggregators}

\begin{algorithm}[H]
    \caption{Routes Splitting Algorithm in DEX Aggregators}\label{Algorithm 4}
\textbf{Problem setting:} \textit{We need to buy the maximal units of token $v_N$ ($\Delta r_n$) with $\epsilon_1$ units of token $v_1$ in a DEX.}

    \hspace*{\algorithmicindent} \textbf{Input}: $\epsilon_1$ units of $v_1$\\
    \hspace*{\algorithmicindent} \textbf{Output}: The maximal units of $v_n$
\begin{algorithmic}[0]    
    \State $K \gets$ the number of splits. 
    \State $v_n$ $\gets$ $0$
    \State $P$ $\gets$ the path set.
    \For{$k$ $<$ $K$}
        \State $(T,TP)$ = the line-graph-based method in \cite{zhang2025line} ($\frac{\epsilon_1}{K}$)
        \State $v_n += T$
        \State $P.add(TP)$
        \State Update the reserve information in the token graph.
    \EndFor
    
    \Return $v_n, P$
\end{algorithmic}
Reorder $P$ to make it an implementable path.
\end{algorithm}

\end{document}